\documentstyle[aps]{revtex}
\begin{document}
\draft
\title{The Balmer-Like Formula for Mass Distribution of Elementary Particle
Resonances}
\author{F.A. Gareev, M.Yu. Barabanov, G.S. Kazacha}
\address{JINR, 141980, Dubna, Moscow Region, Russia}
\date{\today}
\maketitle
\begin{abstract}
Elementary particle resonances have been  systematically    analyzed
using all available experimental data. We have come to the conclusion that
resonance decay product momenta and  masses of resonances are to
be quantized. The Balmer-like formula for mass distribution of elementary
particle resonances was obtained.
These observations
allow us to formulate  a strategy of experimental searches for
new resonances and  systematize the already known.
\end{abstract}
\pacs{PACS numbers: 11., 03.65.-w}

\vspace{1cm}
The merits of
contemporary quark models (the Standard Model) in describing  properties
of elementary
particles are impressive. However, it is well known
that \cite{WEIN97} the only uncertain aspect of the Standard Model is the
mechanism that gives elementary particles their masses. In the simplest
version of the model, these masses depend on constants that specify the
strength of the interaction of  various elementary particles with a new
kind of field but these constants are just
free parameters of the theory. It seems that the mass distribution
of elementary particle resonances is the fundamental problem of modern theory
until we have a final theory of force and matter, particles and fields.
Therefore, the systematic analysis of existing experimental data is
desirable to establish simple phenomenological rules for the mass distribution
of resonances.

One of the most remarkable groundstones for foundation of the quantum theory
was the Balmer formula for the spectra of a hydrogen atom. It
is our purpose here to demonstrate that the Balmer-like formula
for the mass distribution of elementary particle resonances can be obtained
from a systematic analysis  of all available  experimental data.

Some of the resonances have a dominant decay channel, and we suggest that
the momentum of this channel  should manifest itself in properties of
decays through other channels. For example,
it is known that the pion $\pi^{\pm}$ decays through the muon and  neutrino
with a probability about unity and asymptotic momentum $P_{1}=29.7918 MeV/c$.
We are unable to
explain this property of the pion; we only suggest that it is fundamental
for comprehension of the structure of some resonances.
To establish some common properties of the mass distribution of elementary
particle resonances, we check the usefulness of a commensurable principle
\cite{GAR96} of decay  asymptotic momenta. In other words, we check the
following ratio:
$$P_{n}=nP_{1}\;\; n=1,2,3,... \eqno(1)$$
where $P_{1}=29.7918$ Mev/c.

The masses of resonances are calculated by the formula
$$M_{th}=\sqrt{m_{a}^2+P_{n}^2}+\sqrt{m_{b}^2+P_{n}^2}
=\sqrt{m_{a}^2+n^{2}P_{1}^2}+\sqrt{m_{b}^2+n^{2}P_{1}^2},\eqno(2)$$
where $m_{a}$ and $m_{b}$
are the masses of decay products of the resonance to be considered.
The results and corresponding experimental data \cite{PART} are given
in Table \ref{T1}. Note that we present here
the results of calculations for a few resonances only
due to restriction of place. A more complete analysis has been made in \cite{GAS}
containing a few hundred resonances. Our complete analysis containing almost
all data from \cite{PART} can be sent by requirement to everybody.

Of course, there arises a question: what is the role of resonance
decay channels with a very small probability? We know from  nuclear physics
(see, for example, \cite{GAR80}) that basic vectors with small (large)
weights in the
main states of nuclei become the dominant (small) ones in highly excited states.
The same phenomenon is expected to be present in the physics of resonances.
Let us consider as basic
the following decay channel $\pi^{0}\rightarrow \mu^{\pm}e^{\mp}$ with
$P_{1}$=26.1299 MeV/c and fraction $\approx 10^{-8}$\%. The results of
calculations are presented in Table \ref{T2}.

Tables \ref{T1}, \ref{T2} contain rich information, and it is possible to make many
fundamental conclusions based on them:

  We think that the results contained in the tables convincingly demonstrate
the empirical fact of commensurability  of resonance decay product momenta
within the accuracy of existing experimental data. In other words,
resonance decay product momenta are quantized. It is clear from the Tables that
commensurability of momenta does not depend on the type of interaction between
resonance decay products, quantum numbers of resonances and type of particles.
Moreover, commensurability of momenta is justified for all
considered resonances. It seems to be a universal
property of resonances, and many periodic structures are just governed by this
property. A main question arises here: is this commensurability precise or
approximate? We do not know the answer but think that it is as precise  as the
one in solids
and crystals. If so, the common structure of elementary particles must be
analogous to that of solids and crystals.
An excellent possibility for prediction of new resonances and verification
of masses of existing ones arises in any case.

 We are able to interpret some of the resonances as radial excitations.
 For example, $\rho(770)$, $\rho(1450)$ and $\rho(2150)$-mesons
decay through two pions with momenta 358, 719 and 1065 MeV/c, i.e.
their momenta are commensurable.

 The systematics of resonance decay chains and decay products
\cite{PART} in turn suggests a simple idea about the structure of elementary
particle resonances: all heavy particles decay into lighter ones  in a tree-like manner,
and the final decay products are electrons, photons, neutrinos, and protons.
Resonance decay is of a self-similar cascade character and, moreover, the
relevant decay channel momenta are commensurable.

Obviously, the accuracy of resonance masses to be predicted
depends on the accuracy of decay product masses used in calculations.
Therefore, our predictions of resonance masses are to be considered as
preliminary
ones. We ask readers to send  information about more accurate contemporary
experimental resonance mass data. For this we will be indebted.
The obtained results coincided with existing experimental data within the
accuracy of decay product masses.

We have established that experimental values of masses of
$PP$ and $\overline{P}P$ resonances \cite{GAR} coincide within
the accuracy of experimental data.
This remarkable observation was highlighted in our earlier works
\cite{GAR96G,GAR97}. But the systematic analysis of
experimental data brings us to the more general conclusion. The masses of
resonances coincide within the accuracy of experimental data
decaying through the same channels: particle-particle, particle-antiparticle
and antiparticle-antiparticle. This observation is valid for any resonances
independent of the type of  particles, interaction and so on.
If so,  good perspectives for
a combined systematic analysis of this type of resonances are opened.

So, we come to the conclusion that resonances have an extremely rich structure.
They are complex wave systems. All the motions of subsystems (constituents)
are commensurable according to the laws of wave mechanics. Thus, there is
observed self-organization of  constituents of  matter between each other and
also with the whole system irrespective of the type of  interaction between
the constituents. Moreover, the constituents of  matter (clusters) are similar
to each other and to the whole system. Actually, the resonances decay in
sequence in a  tree-like  manner forward for lighter ones so that
the  branches of the tree are similar to each other and to the whole tree.

Our systematic analysis of resonance decay products shows that the
corresponding motions of the constituents do not exist independently. The
motions of constituents in one resonance  are consistent with those of
constituents in the others. Each resonance itself plays a three-fold role:
has a complex structure, is included into other resonances, and takes part in
the exchange between components of a subsystem keeping up the unity of the
structure. Thus, a harmonic unity of motions of all particles is established.

To understand how an unstable particle (resonance) can communicate with some
open channels we have found that it is necessary to develop an analogy  between the
behavior of unstable particles and resonant cavities such as organ pipes and
electromagnetic resonators. Chew,  Gell-Mann, and Rosenfeld were the first who
paid attention to this analogy in 1964 \cite{CHEW64}.
They discussed in detail the possibility of representing  elementary
particle resonances as resonant cavities in full analogy with our approach.
But the 1964  paper  \cite{CHEW64} remained unnoticed.

As the final conclusion we can say that we have established the Balmer-like
formula for masses of elementary particle resonances which is supported
by a systematic analysis of experimental data.

The authors would like to thank the Russian Foundation for Basic Research
for financial support of our work.



%
%

%
%

\newpage
\begin{center}
\begin{table}
\caption{
Invariant masses of resonances decaying through binary channels
with momenta  $P_{n}=n*29.7918$ MeV/c.
}
\label{T1}

\begin{tabular}{cccccccc}
resonances & decay channels&{\bf $P_{exp}$ } &n&{\bf $P_{exp}/n$ }& {\bf $ M_{exp}$} & {\bf $ M_{th}$}&${\bf \mid M_{th}-M_{exp} \mid}$ \\\hline
$\pi^{\pm}$&$\mu^{\pm}\nu_{\mu}$&29.79&1&29.79&139.56995&139.56995&---\\
$\rho(770)$  &$\pi^{\pm}\pi^{\mp}$ &358 &12&29.83  &$768.5\pm 0.6$&767.56&0.94\\
$\rho(1450)$ &$\pi^{\pm}\pi^{\mp}$ &719&24  &29.96  &$1465\pm 25$&1457.00&8.00\\
$f_{2}(1810)$&$\pi^{\pm}\pi^{\mp}$ &896.70&30&29.89&$1815\pm 12$&1809.17&2.17\\
$\rho(2150)$ &$\pi^{\pm}\pi^{\mp}$ &1075.99&36 &29.89&$\sim 2170$& 2163.10&6.90\\
$\hat{\rho}(1405)$&$\eta'\pi^{0}$&358.71&12&29.89&$1406\pm 20$&1404.45&1.55\\
$\hat{\rho}(1405)$&$\eta\pi^{0}$&538.52&18&29.92&$1323.1\pm 4.6$&1320.08&3.02\\
$\rho(1450)$&$\phi\pi^{0}$ & 359.54 &12&29.96&$1465\pm 25$&1462.42&2.58\\
$\rho_{3}(1690)$&$\phi\pi^{0}$& 526.59 &18&29.26&$1691\pm 5$&1704.83&3.83\\
$\rho_{3}(1690)$&$\rho^{0}\rho^{0}$& 352.53 &12&26.88&$1691\pm 5$&1693.46&2.46\\
$\rho(1700)$&$\rho^{0}\rho^{0}$& 363.19 &12&30.27&$1700\pm 20$&1693.46&6.54\\
$\rho(1700)$&$\rho\eta$&534.34&18&29.69&$1701\pm 15$&1703.44&2.44\\
$\rho(2150)$ &$p\overline{p}$&529.77&18&29.43&$2155\pm 15$&2161.41&6.41\\
$\rho_{5}(2350)$&$p\overline{p}$&714.75&24&29.87&$2359\pm 2$&2359.31&0.31\\
$a_{1}(1260)$&$\rho^{0}\pi^{+}$&356.49&12&29.71&$1230\pm 40$&1230.51&0.51\\
$f_{1}(1285)$&$a_{0}(980)\pi^{0}$&235.01&8&29.38&$1282.2\pm 0.7$&1285.87&3.67  \\
$f_{1}(1420)$&$a_{0}(980)\pi ^{0}$&356.10&12&29.68&$1426.8\pm 2.3$&1428.59&1.79\\
$f_{1}(1510)$&$a_{0}(980)\pi^{0}$&421.05&14&30.08&$1512\pm 4$&1506.67&5.33\\
$f_{0}(1370)$&2$K_{S}^{0}$&536.11&18&29.78&$1463\pm 9$&1463.9&0.90\\
$f_{J}(1710)$&2$K_{S}^{0}$&714.82&24&29.78&$1742\pm 15$&1742.31&0.31\\
$a_{2}(1320)$&$\eta\pi^{0}$&535.62&18&29.76&$1318.1\pm 0.7$&1319.31&1.21\\
$X(1^{-+})$&$\eta\pi{-}$&537.84&18&29.88&$1323.1\pm 4.6$\cite{KEK}&1320.45&2.65\\
$X(1775)$&$\rho^{\pm}\pi^{\mp}$   &714&24     &29.75   &$1776\pm 13$&1778.18&2.18\\
$D^{\pm}$&$\rho^{\pm}\eta'$&355&12&29.58&$1869.3\pm 0.5$&1869.9&0.6\\
$\omega_{3}(1670)$&$b_{1}(1235)\pi^{0}$&359&12&29.92&$1667\pm 4$&1663.99&3.01\\
$f_{J}(1710)$&$\rho\rho$& 359.67&12 &29.97&$1697\pm 4$&1695.17&1.83\\
$K_{3}^{*}(1770)$&$K^{0}\eta$&715&24&29.79&$1770\pm 10$&1771.67&1.67\\
$D_{1}(2420)^{0}$&$D^{*}(2010)^{\pm}\pi^{\mp}$&355&12&29.58&$2422.2\pm 1.8$&2425.33&3.13\\
$K_{2}^{*}(1460)^{\pm}$&$K^{\pm}\rho^{0}$&360.36&12&30.03&$\sim 1460$&1456.26&3.74\\
$\rho_{5}(2350)$&$p\overline{p}$&714.75&24&29.87&$2359\pm 2$&2359.31&0.31\\
X(2850)&$p\overline{p}$&1072.51&36&29.79&$2850\pm 5$&2850.00&0.00\\
$\Sigma(1620)$&$\Sigma(1385)^{0}\pi^{0}$&147.77&6&29.58&$1618\pm 3$&1619.19&1.19\\
$\Sigma(1940)$&$\Lambda(1520)^{0}\pi^{0}$&354.82&12&29.57&$1940\pm 20$&1943.12&3.12\\
$\Sigma(1480)$&$p\overline{K}^{0}$&171.08&6&28.51&$1480$&1483.95&3.95\\
$\Lambda(1600)$&$n\overline{K}^{0}$&356.38&12&29.70&$1617$&1618.05&1.05\\
$\Sigma(1385)^{0}$&$\Sigma^{0}\gamma$&177.95&6&29.66&$1383.7\pm 1$&1384.62&0.92\\
$\phi(1020)$&$\eta'(958)\gamma$&59.78&2&29.89&$1019.413\pm 0.008$&1019.20&0.21\\
$\phi(1680)$&$\eta'(958)\gamma$&566.99&19&29.84&$1680\pm 20$&1678.58&1.42\\
$\phi_{3}(1850)$&$\eta'(958)\gamma$&679.61&23&29.55&$1854\pm 7$&1862.85&8.85\\
$J/\psi(1S)$&$\eta'(958)\gamma$&1400.34&47&29.79&$3096.88\pm 0.04$&3096.66&0.22\\
$\phi(4040)$&$\eta'(958)\gamma$&1906.47&64&29.79&$4040\pm 10$&4040.39&0.39\\
$\phi(4160)$&$\eta'(958)\gamma$&1969.22&66&29.84&$4159\pm 20$&4153.38&5.62\\
$\Upsilon(4S)$&$\eta'(958)\gamma$&5246.65&176&29.81&$10580\pm 3.5$&10573.47&6.53\\
$\Upsilon(10680)$&$\eta'(958)\gamma$&5390.29&181&29.78&$10865\pm 8$&10869.02&4.02\\
$\pi_{2}(1670)$&$\mu^{\pm}\nu_{\mu}$&831.66&28&29.70&$1670\pm 20$&1675.00&5.00\\
$\pi(1800)$&$\mu^{\pm}\nu_{\mu}$&894.39&30&29.81&$1795\pm 10$&1793.73&1.27    \\
$\pi_{2}(2100)$&$\mu^{\pm}\nu_{\mu}$&1042.33&35&29.78&$2090\pm 29$&2090.76&0.76\\
$\Sigma(2455)$&$\Lambda^{+}_{c}\pi{-}$&93.82&3&31.27&$\approx 2455$&2452.38&2.62\\
$\Sigma_{c}(2455)^{++}$&$\Lambda^{+}_{c}\pi{+}$&90.27&3&30.09&$2452.9\pm 0.6$&2452.38&0.52\\
$\Xi^{0}_{c}$&$\Lambda^{+}_{c}\pi{-}$&117.55&4&29.39&$2470.4\pm 1.8$&2471.53&1.13\\
$\Sigma(2620)$&$\Lambda^{+}_{c}\pi{-}$&204.91&7&29.27&$2542\pm 22$&2545.33&3.33\\
\hline
\end{tabular}
\end{table}
\end{center}

\begin{center}
\begin{table}
\caption{
Invariant masses of higher excited resonances decaying through binary channels
with  momenta  $P_{n}=n*26.1299$ MeV/c.
}
\label{T2}

\begin{tabular}{cccccccc}
resonances & decay channels&{\bf $P_{exp}$ } &n&{\bf $P_{exp}/n$ }& {\bf $ M_{exp}$} & {\bf $ M_{th}$}&${\bf \mid M_{th}-M_{exp} \mid}$  \\\hline
$\pi^{0}$&$\mu^{\pm}e^{\mp}$&26.1299&1&26.1299&134.9764&134.9764&---\\
$D^{*}_{2}(2460)^{0}$&$\mu^{\pm}e^{\mp}$&1227.18&47&26.11&$2458.9\pm 2.0$&2460.7474&1.85\\
$B^{*}_{J}(5732)$&$\mu^{\pm}e^{\mp}$&2848.02&109&26.13&$5698\pm 12$&5698.2774&0.28\\
$B^{*}_{sJ}(5850)$&$\mu^{\pm}e^{\mp}$&2925.55&112&26.12&$5853\pm 13$&5855.0043&2.00\\
$\Upsilon(1S)$&$\mu^{\pm}e^{\mp}$&4729.59&181&26.13&$9460.37\pm 0.21$&9460.2039&0.17\\
$\omega(782)$&$e^{-}e^{+}$&390.97&15&26.06&$781.94\pm 0.12$&783.8983&1.96\\
$X(1097)$&$e^{-}e^{+}$&548.5&21&26.12&$1097^{+16}_{-19}$&1097.4572&0.46\\
$K^{*}(1410)$&$e^{\pm}\nu_{e}$&706.00&27&26.15&$1412\pm 12$&1411.0160&0.98\\
$\rho(1450)$&$e^{-}e^{+}$&732.5&28&26.16&$1465\pm 25$&1463.2760&1.72\\
$f_{2}(1565)$&$e^{-}e^{+}$&782.5&30&26.08&$1565\pm 20$&1567.7957&2.80\\
$\tau$&$e^{\pm}\gamma$&888.5&34&26.13&$1777^{+0.30}_{-0.27}$&1776.8349&0.17\\
$X(1830)$&$e^{-}e^{+}$&915.00&35&26.14&$\sim 1830$&1829.0949&0.91\\
$\pi_{2}(2090)$&$e^{-}e^{+}$&1045.00&40&26.13&$2090\pm 29$&2090.3940&0.39\\
$f_{0}(2200)$&$e^{-}e^{+}$&1098.50&42&26.15&$2197\pm 17$&2194.9137&2.09\\
$K_{2}(2250)$&$e^{\pm}\nu_{e}$&1123.50&43&26.13&$2247\pm 17$&2247.1735&0.17\\
$D^{*}_{2}(2460)^{0}$&$e^{-}e^{+}$&1229.45&47&26.16&$2458.9\pm 2$&2456.2129&2.69\\
$D^{*}_{2}(2460)^{\pm}$&$e^{\pm}\nu_{e}$&1229.50&47&26.16&$2459\pm 4$&2456.2128&2.79\\
$f_{2}(2510)$&$e^{-}e^{+}$&1255&48&26.15&$2510\pm 30$&2508.4728&1.53\\
$\eta_{c}(1S)$&$e^{-}e^{+}$&1489.90&57&26.14&$2979.8\pm 2.1$&2978.8114&0.99\\
$B^{0}$&$e^{-}e^{+}$&2639.6&101&26.13&$5279.2\pm 1.8$&5278.2449&0.96\\
$B^{\pm}$&$e^{\pm}\nu_{e}$&2639.45&101&26.13&$5278.9\pm 1.8$&5278.2444&0.64\\
$B^{*}_{J}(5732)$&$e^{-}e^{+}$&2849.00&109&26.14&$5698\pm 12$&5696.3232&1.68\\
$B^{*}_{J}(5732)$&$e^{\pm}\nu_{e}$&2849.00&109&26.14&$5698\pm 12$&5696.3232&1.68\\
$B^{*}_{sJ}(5850)$&$e^{-}e^{+}$&2926.5&112&26.13&$5853\pm 15$&5853.1028&.10\\
$B^{*}_{sJ}(5850)$&$e^{\pm}\nu_{e}$&2926.5&112&26.13&$5853\pm 13$&5853.1027&0.10\\
$\Upsilon(10860)$&$e^{-}e^{+}$&5432.5&208&26.12&$10865\pm 8$&10870.0479&5.04\\
$f_{1}(1285)$&$\phi\gamma$&236&9&26.22&1282.2$\pm0.7$&1281.36&0.84\\
$f_{1}(1285)$&$a_{0}(980)\pi^{0}$&234&9&26&1282.2$\pm0.7$&1282.38&0.18\\
$a_{2}(1320)$&$\eta^{\prime}(958)\pi^{0}$&287&11&26.09&1318.1$\pm0.7$&1317.51&0.59\\
$K^{0}_{S}$&$\pi^{0}\pi^{0}$&209&8&26.13&497.672$\pm0.031$&497.66&0.01\\
$K^{*}_{2}(1430)^{\pm}$&$K^{\pm}\gamma$&627&24&26.13&1425.4$\pm1.3$&1425.24&0.16\\
$D^{\pm}$&$\overline{K}^{0}\pi^{\pm}$&862&33&26.12&1869.3$\pm0.5$&1869.11&0.19\\
$D^{0}$&$\overline{K}^{0}f_{0}(980)$&549&21&26.14&1864.5$\pm0.5$&1863.97&0.53\\
$D^{0}$&$\phi\rho^{0}$&260&10&26&1864.5$\pm0.5$&1864.08&0.42\\
$D^{\pm}_{s}$&$f_{0}(980)\pi^{\pm}$&732&28&26.14&1968.5$\pm0.6$&1967.82&0.68\\
$D^{\pm}_{s}$&$\eta^{\prime}(958)\rho^{\pm}$&470&18&26.11&1968.5$\pm0.6$&1968.03&0.47\\
$D_{s1}(2536)^{\pm}$&$D^{0}K^{\pm}$&392&15&26.13&2535.35$\pm0.34$&2535.6&0.25\\
$B^{\pm}$&$e^{\pm}\gamma$&2639&101&26.13&5278.9$\pm1.8$&5278.24&0.66\\
$B^{0}$&$e^{\pm}\mu^{\mp}$&2639&101&26.13&5279.2$\pm1.8$&5280.35&1.15\\
$\eta_{c}(1S)$&$K^{*}(892)^{\pm}\overline{K}^{\mp}$&1307&50&26.14&2979.8$\pm2.1$&2978.38&0.42\\
$\psi(2S)$&$\gamma\chi_{c0}(1P)$&261&10&26.1&3686.00$\pm0.09$&3686.38&0.38\\
$\chi_{b0}(1P)$&$\gamma\Upsilon(1S)$&391&15&26.07&9859.8$\pm1.3$&9860.44&0.64\\
$\Upsilon(2S)$&$\gamma\chi_{b1}(1P)$&131&5&26.2&10023.3$\pm0.31$&10023.41&0.11\\
$\Sigma(1385)^{0}$&$\Lambda\pi^{0}$&208&8&26&1383.7$\pm1.0$&1383.93&0.23\\
$\Lambda^{+}_{c}$&$\Xi(1530)^{0}K^{+}$&471&18&26.17&2284.9$\pm0.6$&2284.25&0.65\\
$\Lambda_{c}(2593)^{+}$&$\Lambda^{+}_{c}\pi^{0}$&261&10&26.1&2593.6$\pm1.0$&2593.9&0.3\\
$\Xi^{+}_{c}$&$\Sigma^{+}\overline{K}^{*}(892)^{0}$&653&25&26.12&2465.6$\pm1.4$&2465.89&0.29\\
$\Xi^{0}_{c}$&$\Omega^{-}K^{+}$&522&20&26.1&2470.3$\pm1.8$&2471.11&0.89\\
$\Xi_{c}(2645)$&$\Xi^{+}_{c}\pi^{-}$&107&4&26.75&2643.8$\pm$1.8&2642.18&1.62\\
\hline
\end{tabular}
\end{table}
\end{center}

\end{document}